\documentclass[%
reprint,
 amsmath,amssymb,
 aps,
prb,
]{revtex4-1}

\usepackage{graphicx}% Include figure files
\usepackage{dcolumn}% Align table columns on decimal point
\usepackage{bm}% bold math
\begin{document}
\title{Hybridization-Driven Orthorhombic Lattice Instability in URu$_2$Si$_2$}
\author{T. Yanagisawa$^1$}
\email{tatsuya@phys.sci.hokudai.ac.jp}
\author{S. Mombetsu$^1$}
\author{H. Hidaka$^1$}
\author{H. Amitsuka$^1$}
\author{M. Akatsu$^{2,3}$}
\author{S. Yasin$^2$}
\author{S. Zherlitsyn$^2$}
\author{J. Wosnitza$^{2,4}$}
\author{K. Huang$^5$}
\author{M. Janoschek$^{5}$}
\email{Present Address: Los Alamos National Laboratory}
\author{M. B. Maple$^5$}
\affiliation{$^1$Department of Physics, Hokkaido University, Sapporo 060-0810, Japan}
\affiliation{$^2$Hochfeld-Magnetlabor Dresden, Helmholtz-Zentrum Dresden-Rossendorf, D-01314 Dresden, Germany}
\affiliation{$^3$Department of Physics, Niigata University, Niigata 950-2181, Japan}
\affiliation{$^4$Institut f\"ur Festk\"orperphysik, TU Dresden, D-01062 Dresden, Germany}
\affiliation{$^5$Department of Physics, University of California, San Diego, La Jolla, CA 92093, U.S.A.}
\date{October 1, 2013}

\begin{abstract}
We have measured the elastic constant ($C_{11}$-$C_{12}$)/2 in URu$_2$Si$_2$ by means of high-frequency ultrasonic measurements in pulsed magnetic fields $H \parallel$ [001] up to 61.8 T in a wide temperature range from 1.5 to 116 K. We found a reduction of ($C_{11}$-$C_{12}$)/2 that appears only in the temperature and magnetic field region in which URu$_2$Si$_2$ exhibits a heavy-electron state and hidden-order. This change in ($C_{11}$-$C_{12}$)/2 appears to be a response of the 5$f$-electrons to an orthorhombic and volume conservative strain field $\epsilon_{xx}$-$\epsilon_{yy}$ with $\Gamma_3$-symmetry. This lattice instability is likely related to a symmetry-breaking band instability that arises due to the hybridization of the localized $f$-electrons with the conduction electrons, and is probably linked to the hidden-order parameter of this compound.

\begin{description}
%\item[Keywords] URu$_2$Si$_2$, hidden order, pulsed-magnetic field, elastic constant, ultrasound
\item[PACS numbers]{71.27.+a, 62.20.de, 62.65.+k}
% PACS, 71.27.+a (Heavy-fermion solids-electron states), 62.20.de (Elastic moduli). 62.65.+k (Acoustic Properties-solids), 75.30.Kz (Magnetic phase transitions)
\end{description}
\end{abstract}

\maketitle
The heavy-fermion compound URu$_2$Si$_2$ exhibits a second-order phase transition involving uranium's $5f$-electron state at $T_{\rm o}$ = 17.5 K, and also exhibits unconventional superconductivity at $T_{\rm c} \sim$ 1.4 K. Though the transition at $T_{\rm o}$ shows clear anomalies in several thermodynamic quantities~\cite{Palstra85, Maple86, Schlabitz86}, the order parameter has not been identified microscopically for more than a quarter of a century; thus the unknown transition has been called ``hidden order (HO)" ~\cite{Shah00}.

In high magnetic fields (using static or pulsed magnetic field) for $H \parallel$ [001] at low temperatures, URu$_2$Si$_2$ undergoes three metamagnetic transitions in the range 35 and 39 T with a collapse of the HO phase~\cite{Harrison03, Kim03, Scheerer12}, where the $c$-axis magnetization increases in three steps and reaches a value of $\sim$1.5$\mu_{\rm B}$/U at 60 T and 1.5 K, which is approximately half of the value expected from the localized $5f$-electron state of U$^{3+}$ (3.62$\mu_{\rm B}$) or U$^{4+}$ (3.58$\mu_{\rm B}$). In addition, the electrical resistivity shows an abrupt decrease above 40 T. This suggest that tuning URu$_2$Si$_2$ by means of a magnetic field decreases the hybridization between 5$f$ and conduction electrons, and, therefore, leads to a reduced effective electron mass~\cite{Kim03, Scheerer12}. Given this situation, it is plausible that the hybridized-electron state observed at low-temperature and in low-magnetic-field regions will change to a relatively light Fermi-liquid state when a high magnetic field is applied along the [001] axis as indicated by the disappearance of the heavy band~\cite{Scheerer12}. 
Thus, the collapse of the HO and the drastic change of the $5f$ electronic state appear to be strongly connected, and suggest that the order parameter of the HO phase is veiled in the screening of the localized $5f$ electronic state via their strong hybridization with conduction electrons states.

In contrast, several recent theoretical models predict that the order parameter of the HO phase is of local nature and based on higher-rank electric and magnetic multipoles~\cite{Harima10, Kusunose11, Thalmeier11}. In particular, the $xy$($x^2$-$y^2$) electric hexadecapole model, which has been proposed by Kusunose and Harima and also partly related to the theory of Haue and Kotliar~\cite{Haule09}, suggests that electric quadrupole moments $O_{xy}$(= $J_xJ_y$+$J_yJ_x$) and $O_2^2$(= $J_x^2$-$J_y^2$) are induced by an in-$c$-plane magnetic field or uniaxial stress via Ginzburg-Landau coupling.

Ultrasonic measurements are a powerful tool to probe the existence of such quadrupoles, as they induce a change of the elastic constants when either temperature or magnetic field are varied. Based on a model with localized $f$ electrons, this is generally understood as the quadrupolar susceptibility. In contrast with the hexadecapol model, our recent ultrasonic measurements of the elastic constant ($C_{11}$-$C_{12}$)/2 in pulsed magnetic fields applied parallel to [100] and [110] have demonstrated that there is no in-plane anisotropy up to 68.7 T~\cite{Yanagisawa13}. While higher magnetic fields or higher measurement accuracy will be required to ultimately rule out the existence of the antiferro-hexadecapolar order, this suggests that the HO is not explained by this model unless otherwise considered possible effect of itinerant $f$ electrons.

On the other hand, our results for $H \parallel$ [001] show that the change of the elastic constant ($C_{11}$-$C_{12}$)/2 up to $\sim$50 T is comparable to the elastic softening in the temperature dependence from 120 K to $T_{\rm o}$. This indicates that the low-temperature electronic state of URu$_2$Si$_2$ exhibits a lattice instability with $\Gamma_3$ symmetry ($x^2$-$y^2$-type) that is strongly related to the origin of the hybridized electronic state and the HO~\cite{Yanagisawa13}. 

In the present work, we explore a wider temperature range from 1.5 K to 116 K up to 61.8 T with $H \parallel$ [001] for ultrasonic measurements on URu$_2$Si$_2$ in order to check the temperature dependence of the $\Gamma_3$ lattice instability in the high-magnetic-field region (40 $\le H \le$ 61.85 T), where both the HO and hybridization effects are suppressed. Our results demonstrate clear evidence for an orthorhombic lattice instability that appears to be closely tied to the HO of URu$_2$Si$_2$.

\begin{figure}[b] 
\includegraphics[width=1.0\linewidth]{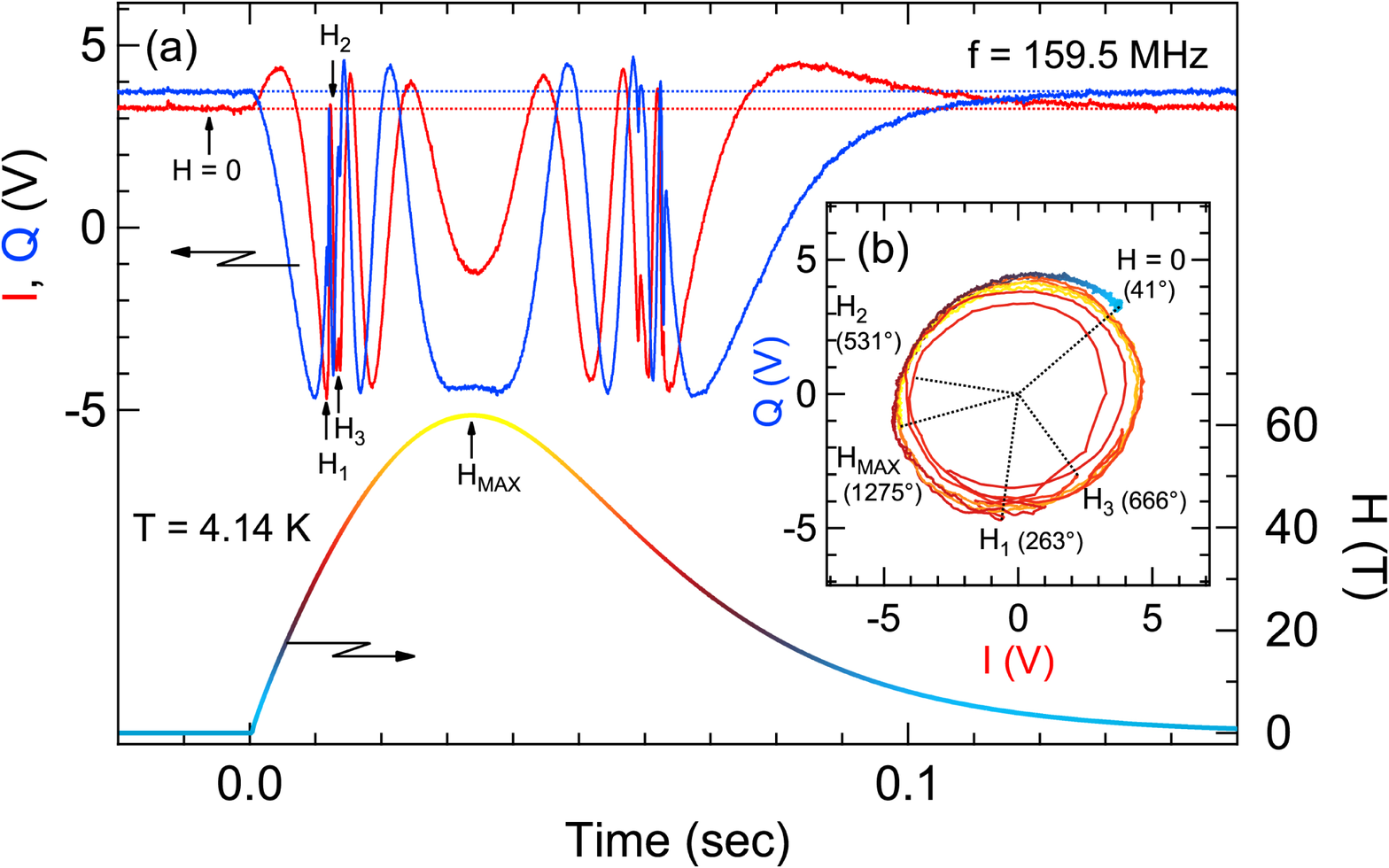}
\caption{(Color online) (a) Magnetic field (right axis) and output voltage of in-phase ($I$) and quadrature ($Q$) signals of the gated integrators at 4.14 K and 159.5 MHz (left axis) as a function of the elapsed time. Inset (b) shows the Lissajous curve of the $I$ and $Q$ signals.}
\end{figure}

A single crystal of URu$_2$Si$_2$ was grown at UC San Diego using the Czochralski method, and subsequently cut and mirror-like finishing, resulting in a length of 3.8024 mm along the [110] axis. Ultrasonic measurements were performed at the Hochfeld-Magnetlabor Dresden using the quadrature procedure described in more detail in Ref. \onlinecite{Yanagisawa13}. For the present measurements, long-pulsed magnetic fields with a duration of $\sim$150 ms and a maximum field of 61.85 T have been used. Long pulses offer the advantage that heating of the sample due to eddy currents, as well as adiabatic cooling and heating frequently observed in the vicinity of phase boundaries, when fast pulses are employed, are insignificant. 

Figure 1 shows in-phase and quadrature signals of an ultrasonic wave with a frequency of 159.5 MHz for the present URu$_2$Si$_2$ sample as well as the magnitude of the pulsed magnetic field as a function of elapsed time.
The arrow denoted by $H_{\rm MAX}$ indicates the time to reach the maximum field. In addition, $H_1$, $H_2$, and $H_3$ denote three phase boundaries observed in the high-magnetic-field phase diagram of URu$_2$Si$_2$~\cite{Mydosh12}. Notably, $H_1$ marks the HO transition, and $H_2$ and $H_3$ are the entry in and exit out of the so-called phase III~\cite{Mydosh12}, which is discussed as possible reentrant HO~\cite{Harrison03}. The concentric Lissajous curve (shown in the inset of Fig. 1) demonstrates that the wave-detection system maintains good stability for the output levels and phase shift, and, in addition, that ultrasonic data obtained during magnetic-field sweeps represents true isotherms below 4.14 K.

\begin{figure}[t] 
\includegraphics[width=0.9\linewidth]{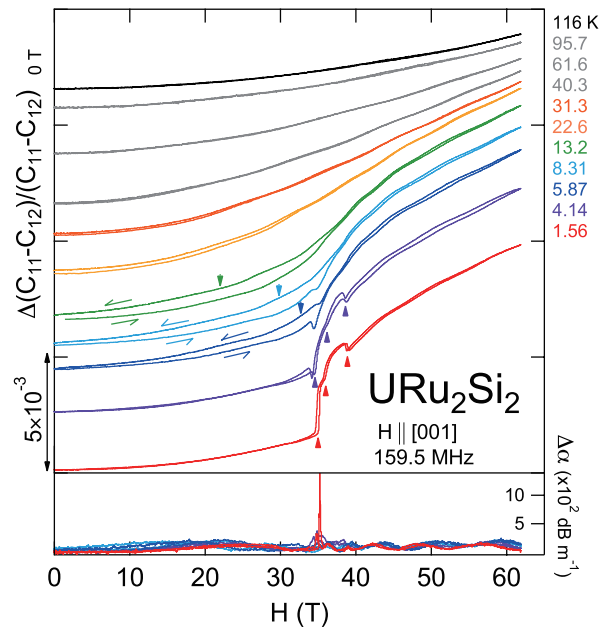}
\caption{(Color online) Upper panel: Elastic constant ($C_{11}$-$C_{12}$)/2 normalized by the values at 0 T vs. the magnetic field $H \parallel$ [001] of URu$_2$Si$_2$ at various fixed temperatures between 1.56 and 116 K. Lower panel: Sound-attenuation change $\Delta \alpha$ vs. $H$ for 1.56 to 13.2 K. All the data were taken with both up- and down-sweeps of the magnetic field, as indicated by arrows.
}
\end{figure} 

Observed isotherms of the normalized elastic constant ($C_{11}$-$C_{12}$)/2 are shown in Fig. 2 as a function of magnetic field $H \parallel$ [001]. Both curves for rising and falling magnetic fields are plotted and have been shifted vertically for clarity. The three phase transitions $H_1$, $H_2$, and $H_3$ discussed above are visible in our data as clear elastic anomalies for magnetic fields between 35 and 39 T and temperatures below 5.87 K as denoted by the upward facing triangles. Intriguingly, the anomaly $H_1$ at $\sim$35 T and 1.56 K is marked by a sudden upturn, in contrast to the relatively small upturn with a precursor small dip observed at 4.14 and 5.87 K. We note that such differences are most likely due to the complex nature of the phase diagram in the vicinity of the upper phase boundary of the HO and related to the novel phases II or V reported in Refs. \cite{Harrison03, Kim03, Scheerer12, Suslov03, Mydosh12}. 

As apparent from Fig. 2, the magnetic field dependence of ($C_{11}$-$C_{12}$)/2 in URu$_2$Si$_2$ looks very similar to magnetization vs. $H$ (applied along the $c$-axis)~\cite{Scheerer12}, and thus is reminiscent of magneto-elastic coupling. We can estimate the effect of the magneto-striction on ($C_{11}$-$C_{12}$)/2 using the recently reported thermal-expansion data obtained in static magnetic fields up to 45 T~\cite{Mydosh12}. Using the Ehrenfest relation~\cite{Knafo07}, and the pressure dependence of the $c$-axis at $H_1$~\cite{Inoue01}, we find that the change of volume at $H_1$ is of the order of $\sim 10^{-4}$. The change of the [110] axis will be less or of the same order, and accordingly ($C_{11}$-$C_{12}$)/2 will be affected with a factor of 2 in the phase comparative method. In turn, we conclude that the dramatic change of $\Delta$($C_{11}$-$C_{12}$)/($C_{11}$-$C_{12}$) of $6 \times 10^{-3}$ observed in our measurements near $H_1$ to $H_3$ is an order of magnitude larger than the estimated influence of the magneto-striction, and, therefore, suggests the presence of an additional effect. For completeness, we note that our argument is only complete in the case where the CEF effect of the quadrupole moment $O_2^2$ is also negligible, {\it e.g.,} not in the case of the $\Gamma_5$ doublet ground state. On the other hand, clear hystereses is observed below $\sim$38 T for 5.87, 8.31, and 13.2 K. The hysteresis suggests that we do not have isothermal conditions, at least in this temperature range during up and down sweeps, probably due to strong magneto-caloric effects of this non-Fermi-liquid region relative to other temperature and magnetic field regions. Indeed, these regions may correspond to soft quantum fluctuations arising in the vicinity of the quantum critical end point associated with the destruction of the HO as suggested in Ref. \onlinecite{Garst05}.

\begin{figure}[t] 
\includegraphics[width=0.9\linewidth]{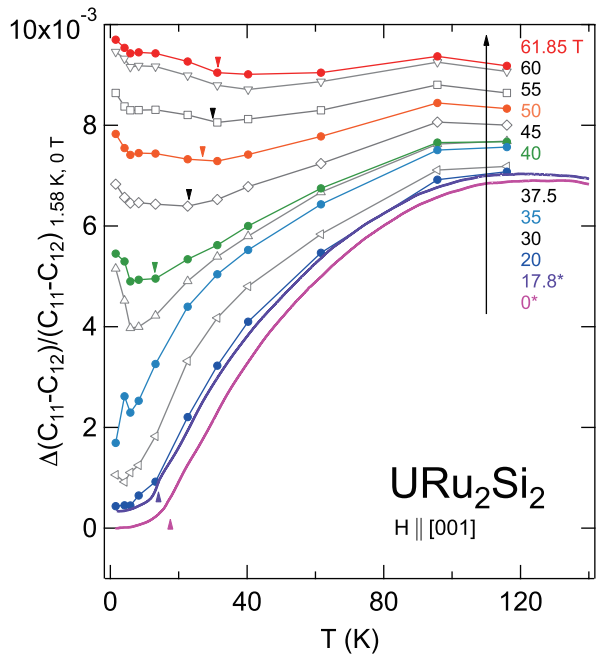}
\caption{(Color online) Normalized elastic constant ($C_{11}$-$C_{12}$)/2 vs. temperature at various magnetic fields $H \parallel$ [001] from 0 to 61.85 T. The curves indicated by a magnetic field with an asterisk, 0 T and 17.8 T, have been measured with the conventional method as a basis for the conversion from isotherms to curves at constant fields. Upward and downward facing triangles indicate the HO transition and local minimum, respectively.
}
\end{figure}   

Figure 3 represents the elastic constant change vs. temperature at several fixed magnetic fields, which was obtained by converting isotherms of ($C_{11}$-$C_{12}$)/2 recorded as a function of magnetic field (Fig. 2). Here, the data sets at zero magnetic field and 17.8 T were measured as function of temperature at constant fields using a standard phase-comparator method and served as a basis to extract the temperature dependence of ($C_{11}$-$C_{12}$)/2 from Fig. 2. The softening of 7$\times$10$^{-3}$ in the relative change of ($C_{11}$-$C_{12}$)/2 below 120 K at zero magnetic field is gradually suppressed with increasing magnetic fields. Above 40 T, ($C_{11}$-$C_{12}$)/2 shows a minimum, indicated by downward facing triangles, which shifts to higher temperatures with increasing magnetic field.

Figures 2 and 3 are compiled into a three-dimensional plot in Fig. 4. It is clearly seen that the softening of ($C_{11}$-$C_{12}$)/2, {\it i.e.,} the $\Gamma_3$ lattice instability, is enhanced in the red-colored region in and around the HO phase, where the strong hybridization is also developed. Conversely, the lattice instability disappears at high temperatures and high magnetic fields, where the HO collapses, as expected from previous work~\cite{Scheerer12}. Our present results obtained over a wide temperature and magnetic-field range, therefore, strongly support that the view that the hybridized electronic state entails the $\Gamma_3$ lattice instability.

\begin{figure}[b] 
\includegraphics[width=1.0\linewidth]{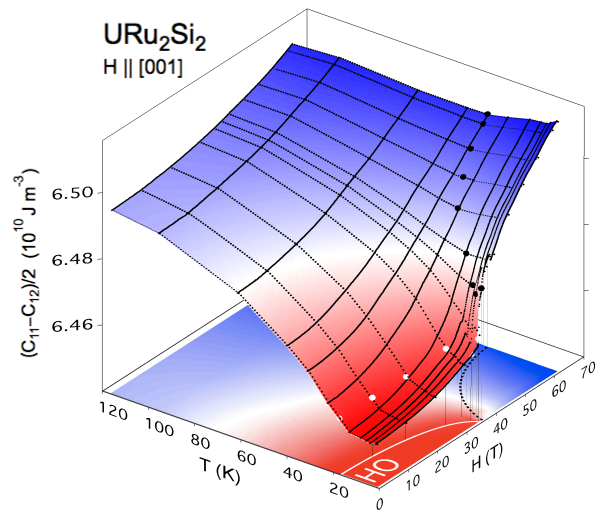}
\caption{(Color online) Three-dimensional plots of the elastic constant ($C_{11}$-$C_{12}$)/2 vs. temperature and magnetic field along the $c$ axis of URu$_2$Si$_2$. The white and black circles indicate the hidden-order transition and local minima found in the temperature dependence of ($C_{11}$-$C_{12}$)/2, respectively. The bottom of the box shows the magnetic field-temperature phase diagram of URu$_2$Si$_2$ (See text for details).
}
\end{figure}

The open (white) and full (black) circles on the surface plot in Fig. 4 indicate the hidden-order transition defined by other experiments and local minima found in the temperature dependence of ($C_{11}$-$C_{12}$)/2, respectively. The dotted curve and the white line are projections of the position of the local minima on the $H$-$T$ phase diagram illustrated onto the basal plane of the plot. We note that the $H$-$T$ phase diagram obtained in this work is similar to the phase diagram presented in Ref. \onlinecite{Scheerer12} that was constructed via high-magnetic-field measurements of the electrical resistivity and magnetization.

Above 40 T, where the hybridization of $f$- and conduction electrons is reduced, we are able to reproduce the temperature and magnetic field dependence of ($C_{11}$-$C_{12}$)/2 employing the theory for a quadrupolar susceptibility based on a localized 5$f$ electron with appropriate CEF scheme. By using the following CEF scheme, $\Gamma_1^{(1)}$-$\Gamma_2$(60 K)-$\Gamma_3$ (178 K), which we have also employed in our previous work~\cite{Yanagisawa13}, the features of ($C_{11}$-$C_{12}$)/2 are qualitatively explained with the exception of the upturn below the local minimum. Since the position of the local minimum in the temperature dependence of the elastic constant (as shown in Fig. 3) may depend on multiple sources such as the gradient of the phonon-background, and the onset of the enhanced magnetization, the reason for this upturn is currently unclear. Further measurements in pulsed magnetic fields using different ultrasonic modes, and magnetic-field directions will help to understand the details of the temperature dependence of ($C_{11}$-$C_{12}$)/2 at the high-magnetic-field region. In particular, performing measurements on the non-magnetic reference sample ThRu$_2$Si$_2$ to obtain the phonon background will facilitate this analysis.
 
Next, we consider the possible origin of the $\Gamma_3$-lattice instability that we observe below 40 T. We expect this orthorhombic instability of URu$_2$Si$_2$ is caused by hybridization of the 5$f$ electrons with the conduction electrons, since the non-$5f$ reference compound ThRu$_2$Si$_2$ does not show a softening in the ($C_{11}$-$C_{12}$)/2 mode at zero magnetic field~\cite{Yanagisawa12}. In addition, the softening appears to be sensitive to chemically induced pressure and/or carrier doping as a small amount of Rh substitution (7\%) significantly reduces the magnitude of the softening of ($C_{11}$-$C_{12}$)/2 below 120 K~\cite{Yanagisawa14}. This suggests that for the magnetic-field region below 40 T, the so-called band Jahn-Teller (BJT) effect would be an explanation for the observed elastic response in contrast to the localized picture that we used at high magnetic fields. Here, the energy gain achieved through the formation of a hybridized band leads to this Jahn-Teller-type deformation.

The effect of the BJT distortion on the elastic constants was first discussed in the ultrasonic study of CeAg, LaAg, and their alloys with In~\cite{Niksch87, Knorr80}. In the $Ln$Ag ($Ln$ = La, Ce) compounds, the deformation-potential coupling effect  was confirmed by band-structure calculations and structural analysis of microscopic measurements. At present, microscopic measurements, such as  $^{29}$Si-NMR or high-precision x-ray and neutron scattering~\cite{Takagi07, Amitsuka10, Tabata14}, provide no evidence for a structural symmetry breaking in URu$_2$Si$_2$. Since the elastic softening of URu$_2$Si$_2$ ($\sim$ 0.7\%) is much smaller compared to those of the BJT compounds ($\sim$15-50\%), we expect that a putative band deformation would be of a staggered type, instead of the uniform type observed in the BJT compounds~\cite{Niksch87, Yoshizawa12}. There is, however, no standard formulation of the BJT effect for such a staggered potential deformation. Such a Fermi-surface instability ({\it i.e.,} evoked by Fermi-surface nesting along the $a$ axis with $Q \ne$ 0, predicted theoretically by Oppeneer {\it et al.}~\cite{Oppeneer11} and confirmed experimentally by Kawasaki {\it et al.}~\cite{Kawasaki11}) has an indirect effect on physical quantities at the Brillouin-zone center ($Q=0$) for bulk measurements, in general. Thus, exotic effects such as a higher-order coupling or mode-mode-coupling of two different fluctuation mechanisms may need to be considered to interpret the present results.

We now consider differences between our result, and the theoretical explanation of recent magnetic-torque measurements~\cite{Okazaki11}, which have been interpreted as spontaneous 4-fold rotational symmetry breaking in the tetragonal basal plane that takes place in the HO phase. Here,  Thalmeier {\it et al.} and Ikeda {\it et al.} have proposed $\Gamma_5^+$ (E$^+$)-type quadrupole and $\Gamma_5^-$ (E$^-$)-type dotriacontapole order parameters, respectively, to explain the torque-measurement results (here the sign of $+$ and $-$ indicate the parity of the time-reversal symmetry)~\cite{Thalmeier11, Ikeda12}. Here, the elastic constant $C_{44}$ corresponds to E$^+$ ($\Gamma_5^+$)-symmetry and is expected to show characteristic softening above $T_{\rm o}$ and a relatively large anomaly at $T_{\rm o}$ compared to that of ($C_{11}$-$C_{12}$)/2 with B$_{2g}$ ($\Gamma_3$)-symmetry via electron-phonon interaction, even if the order parameter exhibits nematicity~\cite{Fujimoto11}. However, no distinct change of $C_{44}$ has been observed at around $T_{\rm o}$~\cite{Kuwahara97}, except for a tiny kink, which could be caused by a thermal expansion effect on the sound velocity measurement, and thus, the $\Gamma_3$ lattice instability is highlighted.
%Here,  the elastic constants $C_{44}$ or $C_{66}$ are expected to show relatively large anomalies at $T_{\rm o}$ via electron-phonon interaction, due to the corresponding local lattice distortions of ($yz$, $zx$)-type ($\Gamma_5$) or $xy$-type ($\Gamma_4$), respectively, even if they additionally exhibit nematicity~\cite{Fujimoto11}. However, to date, no distinct change of $C_{44}$ and $C_{66}$ has been observed at $T_{\rm o}$~\cite{Kuwahara97}, except for a tiny kink at $T_{\rm o}$, which could be caused by a thermal expansion effect on the sound-velocity measurement.
One simple interpretation for these discrepancies may be the anisotropic Jahn-Teller-type coupling between lattice and the electron systems, {\it i.e.,} the electron-phonon coupling constant could be `extremely' weak for transverse modes but finite only for the ($C_{11}$-$C_{12}$)/2 mode. Otherwise, we have to simply conclude in agreement with NMR and x-ray diffraction~\cite{Takagi07, Tabata14} that there is no lattice symmetry breaking of $\Gamma_5$ and $\Gamma_4$ symmetry in the HO phase.

It has also been pointed out that the formation of micron-size domains may wash out measurements of bulk properties in URu$_2$Si$_2$~\cite{Okazaki11}. However, in case the continuum approximation is effective, that is when the ultrasonic wavelength in the solid sample is much longer than a force range due to the internal stresses, which occurs when a body is deformed by the elastic-wave propagation~\cite{Luethi}, and also the wavelength is shorter than the domain size or at least comparable with that, ultrasonic measurements will not be influenced by the existence of domains. For the present experiment this condition is fulfilled as the wavelength of the ultrasound waves is estimated $\lambda \sim$ 11-15 $\mu$m for $v \sim$ 2511 ms$^{-1}$ and $f$ = 225-159 MHz. Moreover, no unnecessary reflection or absorption of the ultrasound has been observed, and ultrasonic echoes have been well separated. %, which also implies that the present measurement on the single crystal is less influenced by the possible domain walls
On the other hand, a recent report of thermal-expansion measurements~\cite{Kambe12} suggests that multiple domains may orient to form an orthorhombic mono-domain state when small in-plane uniaxial stress is applied. Therefore, the effects of uniaxial pressure on the elastic response will need to be probed to exclude any influence from domains entirely.

In conclusion, we have investigated the lattice instability existing within the HO and the associated hybridized electron state in URu$_2$Si$_2$. Our results demonstrate that the orthorhombic instability related to the electronic hybridization arising in the HO state $x^2$-$y^2$-type ($\Gamma_3$) symmetry in stark contrast to the $\Gamma_5$- or $\Gamma_4$-type HO parameters that have been recently proposed on the basis of theory.

The authors would like to thank Dr. William Knafo, Dr. Hiroaki Kusunose, and Dr. Yoichi Ikeda for fruitful discussions and also thank Hitoshi Saitoh and Yuto Watanabe for helping with the preliminary ultrasonic measurements. This work was supported by JSPS KAKENHI Grant No.23740250 and No.23102701, and U.S. DOE. Grant No. DE-FG02-04-ER46105. We acknowledge the support of the HLD at HZDR, member of the European Magnetic Field Laboratory (EMFL). One of the authors, M. J., acknowledges financial support from the Alexander von Humboldt Foundation.

\end{document}